\documentclass[aps,prd,nofootinbib,twocolumn,showpacs]{revtex4}
\usepackage{natbib}
\usepackage{graphicx}
\usepackage{color}
\usepackage{textcase}
\usepackage{amsmath, amsthm, amsfonts}
\usepackage{enumerate}
\usepackage{soul}
\usepackage[normalem]{ulem}
\usepackage[english]{babel}
\usepackage[T1]{fontenc}
\usepackage{amsmath}
\usepackage{hyperref}
\usepackage{multirow}
\usepackage{subfigure}

\definecolor{amber}{rgb}{1.0, 0.75, 0.0}

\begin{document}

%\title{Study of the errors in the reconstruction of the gravitational wave signal from the detector responses}
\title{Estimation of the gravitational wave polarizations from a non-template search}

\author{Irene Di Palma}
\email{Irene.DiPalma@roma1.infn.it}
%\affiliation{Albert-Einstein-Institut, Am M\"uhlenberg 1, 14476 Golm}
\affiliation{Istituto Nazionale di Fisica Nucleare, Sezione di Roma, Italy}
\affiliation{Universit\'{a} di Roma La Sapienza, I-00185 Roma, Italy}
\author{Marco Drago}
\email{Marco.Drago@aei.mpg.de}
\affiliation{Albert-Einstein-Institut, Max-Planck-Institut f\"{u}r Gravitationsphysik, Callinstrasse 38, 30167 Hannover, Germany}
\affiliation{Leibniz Universit\"{a}t Hannover, Hannover, Germany}
%\author{}
%\affiliation{}

%\date{} % Activate to display a given date or no date (if empty),
         % otherwise the current date is printed 

\begin{abstract}
Gravitational wave astronomy is just beginning, after the recent success of the four direct detections of binary black hole (BBH) mergers, the first observation from a binary neutron star inspiral and with the expectation of many more events to come.
%The gravitational wave astronomy is starting, just after the success of the first direct detection of gravitational waves and 
%the future expectations of new events.
%{\color{blue}\emph{The new challenge will be to extract from the waveform shape information about the generating source, which will need a good reconstruction of the signal characteristics from the detector data. QUESTA NON e CHIARA, io la toglierei}}
Given the possibility to detect waves from not perfectly modeled astrophysical processes, it is fundamental to be ready to calculate
the polarization waveforms in the case of searches using non-template algorithms.
In such case, the waveform polarizations are the only quantities that contain direct information about the generating process.
We present the performance of a new valuable tool to estimate the inverse solution of gravitational wave transient signals, starting
from the analysis of the signal properties of a non-template algorithm
that is open to a wider class of gravitational signals not covered by template algorithms.
We highlight the contributions to the wave polarization associated with the detector response, 
the sky localization and the polarization angle of the source.
%Through extensive studies, we first derive the analytic contributions of the different source errors in the reconstruction and then we compare them with the estimations coming from Monte Carlo simulations, considering a network of advanced gravitational wave detectors.
In this paper we present the performances of such method and its implications by using two main classes of transient signals, resembling the limiting case for most simple and complicated morphologies. Performances are encouraging, for the tested waveforms: the correlation between the original and the reconstructed waveforms spans from better than 80\% for simple morphologies to better than 50\% for complicated ones. For a not-template search this results can be considered satisfactory to reconstruct the astrophysical progenitor.
\end{abstract}

\maketitle

\section{Introduction}
%general GW
Gravitational waves (GW) were predicted by Einstein's theory of general relativity, in 1916 \cite{ein, ein1}. GWs are dynamic strains in space-time that travel at the speed of light and are generated by non-axisymmetric acceleration of mass. 
The discovery of the binary pulsar system PSR B1913+16 by Hulse and Taylor \cite{hulse} and subsequent observations of
its energy loss by Taylor and Weisberg \cite{taylor} demonstrated the indirect existence of gravitational waves. Such discoveries led to the identification of the importance of direct observations of gravitational waves to study relativistic systems and test general relativity.
\\
%Detectors
The GW community developed a network of ground-based laser interferometers including the two LIGO and Virgo detectors \cite{LIGO, VIRGO}. 
LIGO built the pair of detectors in Hanford (Washington) and Livingston (Louisiana), while Virgo the one in Pisa (Italy) . These
three detectors already took joint scientific runs from 2007 to 2010 putting interesting upper limits on the 
detections of GWs \cite{S6,S6CBC,S6CW,S6Stoch}. Also in Europe, the smaller GEO600 detector \cite{geo} has been running and 
keeping watch on the GW universe, especially while its larger siblings were down for upgrades.
The construction of the second generation interferometers \cite{AdL,AdV} led to the first observing run in September 2015 for
the Advanced LIGO detectors \cite{Harry}. The Virgo detector
has been upgraded into Advanced Virgo \cite{VC} and recently joined the scientific run with LIGO.
Looking ahead, the KAGRA (Kamioka Gravitational Wave Detector) detector \cite{Kagra, Ken} is under construction in an underground site at the Kamioka mine, in Japan.
Recently, another interferometer has been approved with an Indian location \cite{India}.
\\
%THE EVENT
%The first outstanding result from the two LIGO interferometers has been made 
On September 14, 2015, the two LIGO interferometers detected the merge of a binary black hole system for the first time \cite{event, burstPaper, cbcPaper}. The signal event, named GW150914, simultaneously observed in the two LIGO observatories \cite{cbcPaper} and first alerted by a low-latency analysis for generic gravitational wave transients \cite{cWB2008, cWB2015}, matched the waveform predicted by general relativity of the inspiral and merge of a pair of back holes and the ringdown of the resulting single one \cite{astroevent, bbcevent}. Three other events were detected on December 2015, January 2017 and June 2017, together with a candidate event on October 2015, they also matched waveforms generated by two black holes orbiting around each other and merging in a single one \cite{BoxingDay,GW170104,GW170806,BBH-Mergers}. 
Advanced Virgo became operational on August 1, 2017 to join the second scientific run of the Advanced LIGO detectors. The three-detector network identified gravitational waves from a binary black hole coalescence, GW170814, improving the sky localization of the source, and reducing the area of the 90\% credible region from 1160 $deg^{2}$ using only the two LIGO detectors to 60 $deg^{2}$ using all three detectors \cite{GW170814}. On August 17, 2017 the Advanced LIGO and Advanced Virgo detectors made their first observation of a binary neutron star inspiral \cite{GW170817}. The combination of data from the three interferometers allowed the most precisely localized gravitational-wave signal yet and enabled an extensive electromagnetic follow-up campaign that identified a counterpart near the galaxy NGC 4993, consistent with the localization and distance deduced from gravitational wave data \cite{GWGRB}. Using the association between the luminosity distance directly measured from the gravitational wave signal and the galaxy NGC 4993 it is possible also to infer the Hubble constant \cite{Hubble}.
%Virgo contributed to the detection of other two gravitational waves, \cite{GW170814, GW170817}, especially for the source localization. The last one, in particular, was generated by the fusion of two neutron stars and it was the first GW event associated to electromagnetic counterparts, allowing new joint studies within the GW and EM (like the origin of short gamma ray bursts\cite{GWGRB}or the calculation of the Hubble constant\cite{Hubble}.}}
These results affirm the beginning of GW astronomy as well as provide unprecedented observational insights into the physics of binary black holes \cite{AstrophImplic}, the physics of binary neutron stars and the beginning of the multi-messenger astronomy.\\
%{\color{red} SUBTITLE ?}\\
%\section{}
%- comparison between template search\\
%- burst, lack of model, necessary ot start from detector response.\\
%Promising expectations of new events coming from the universe 
Future events might pose the problem of detecting signals that are not strictly modeled
like these first detected systems. Already aware of this issue, the GW community has developed coherent data analysis techniques
which do not require prior knowledge of the signal and are open to a wide possible set of waveform shapes. These procedures
have been applied in the past analyses, especially in burst searches \cite{S5,S6}. We use the word \emph{burst} to identify all the signals which have
limited time duration (less than seconds) and include also astrophysical processes for which there is not a complete model
of the expected gravitational wave. For this reason, in the burst searches no particular assumptions on the waveform are supposed.
%The analysis is general and applies to arbitrary networks of interferometric gravitational wave detectors.
Such coherent methods \cite{cha, and} combine data from multiple detectors and create a unique list of candidate events for the whole network. A well known advantage of coherence is its utility in rejecting background noise glitches \cite{wen, laura}. Glitch rejection is particularly important since it is the limiting factor in the sensitivity of current burst searches, where a confident detection of a gravitational wave burst depends critically on how many glitches pollute the background estimation. \\
A consistent difference of these methods with respect to modeled search filter algorithms, is that they do not estimate directly the
two waveform polarizations, but they reconstruct the projections of the polarizations for the different detectors, as done in \cite{cWB2015, BW}.
%, like the coherent Waveburst 
%algorithm \cite{cWB2015} and the BayesianWave follow-up algorithm \cite{BW}. These pipelines, for instance, both give 
%independent estimations of the GW150914 event which were consistent among the two estimations \cite{burstPaper}. 
In this paper, we present a fully new algorithm that can serve as followup tool to reconstruct the original waveform polarizations starting from the information given
by a generic un-modeled pipeline. As example, we show the results of the application of this new algorithm to coherent Waveburst, the same pipeline that made the first alert of the GW150914 signal \cite{event}.
%In this paper, we study the possibility to reconstruct the original waveform polarizations starting from the information given by an unmodeled algorithm, which is the same that made the first alert of the GW150914 signal \cite{event}.
In section \ref{Sec: theory} we introduce the theoretical calculations behind this work, and how we calculate the signal polarizations starting from the information given
of the detector response and the source localization. Section \ref{Sec: sim} describes the working
condition of the results presented in section \ref{Sec: results} .
%These algorithms are designed to look for unmodelled gravitational wave bursts, 
%this means that it is the most general approach we can consider but it has also the disadvantage of computational costs to estimate the background due to noise, which requires repeated reanalysis of the data using time shifts. That said, it is critical to have a precise estimations of the different characteristics of a signal and furthermore, be able to compare these values with the theoretical expectetions.
%For this reason, in this paper  we've detailed extensive studies of the errors in the reconstruction of the gravitational wave signal from the detector responses.

\section{Inverse solution}\label{Sec: theory}
The projection of the gravitational wave polarizations on a single interferometer is described by the so-called
antenna patterns, which define
the relative interferometer sensitivity in different directions. Each detector is sensitive to a linear combination of the two 
polarizations and has a quadrupolar antenna pattern. In the notation of \cite{sch}, we consider the interferometer's arms along the 
axes $x$ and $y$, hence a generic gravitational wave can be described by the two polarization components $(h_+(t), h_\times(t))$ 
referred to the $x-y$ plane that is rotated by the polarization angle $\psi$, while the arrival direction is given by the spherical coordinates $\theta$, and  $\phi$ relative to the detector's axes.
For convenience we use the same approach, shown in \cite{Sergey}, to represent detectors as vectors.
The projection on a network of interferometric detectors is defined by the vector of detector responses:
\begin{equation}\label{Eq: net response}
{\bf \xi_\Delta(t)} = {\bf F_+(\theta, \phi, \psi)} h_+(t) + {\bf F_\times(\theta, \phi, \psi)} h_\times(t)
\end{equation}
where each vector component is referred to a specific detector $k$. The quantity 
$\{\xi_\Delta(t)\}_k = \xi(t+\Delta_k)$ takes into account the 
relative difference of arrival time between
the detector location and the Earth center as a reference location ($\Delta_k$, depending on the $\theta, \phi$
coordinates), while ${\bf F_{+,\times}}$ are the detector antenna patterns, that are related to the 
relative orientation of the detector arms with respect to the source direction and the wave polarization frame.
Since the new tool we are presenting uses the outputs of a standard coherent GW transient pipeline, 
i.e. the source direction ($\theta$,$\phi$,$\psi$) and the detector response vector
 (${\bf \xi_\Delta(t)}$), we have to solve the system in Eq. \ref{Eq: net response}  to compute the polarization patterns.

From the literature \cite{PRD2011, Reed} it is known that it is not possible to distinguish among the two polarizations 
with only two detectors, for this reason we consider the case of networks composed of $N>2$ detectors.
In such case, we have for each data sample a redundant number of equations with respect to the unknown variable. 
%{\color{blue} For convenience we use the approach of representing detectors as vectors, like in  }\cite{Sergey}.
%{\color{blue}  Since the new tool we are presenting uses the outputs of a standard coherent GW transient pipeline (by combining their data in the most sensitive way) as the source direction ($\theta$,$\phi$,$\psi$) and the detector response vector} (${\bf \xi_\Delta(t)}$), {\color{blue}we start by solving the system in} Eq.\ref{Eq: net response}  {\color{blue}to compute the polarisation patters of the above quantities.}
By introducing the following scalar products:

%We define the following two scalar products and to consider at the same time the contributions of all detectors:
\begin{equation}\label{Eq: scalar product}
\begin{matrix}
{\bf F_+} \cdot {\bf \xi_\Delta(t)} =  |{\bf F_+}|^2 h_+(t) + {\bf F_+} \cdot {\bf F_\times} h_\times(t)\\
{\bf F_\times} \cdot {\bf \xi_\Delta(t)} = {\bf F_\times} \cdot {\bf F_+} h_+(t) + |{\bf F_\times}|^2 h_\times(t)
\end{matrix}
\end{equation}

we reduce the problem to two equations in two variables ($h_+(t)$,$h_\times(t)$). All the other
quantities are given by the un-modeled algorithm output: ${\bf \xi_\Delta(t)}$ are the detectors responses, while from the 
estimation of the source direction ($\theta$,$\phi$,$\psi$) we can calculate both
the antenna patterns (described by the vectors ${\bf F_{+,\times}}$) and the relative difference of arrival time between detectors ($\Delta_k$). 
The system is easily solvable, in fact applying the Cramer's rule, we obtain:

\begin{equation}\label{Eq: solution}
\left\{
\begin{matrix}
h^{(r)}_+(t) = \frac{({\bf F_+} \cdot {\bf \xi_\Delta(t)})(|{\bf F_\times}|^2)-({\bf F_\times} \cdot {\bf \xi_\Delta(t)})({\bf F_+} \cdot {\bf F_\times})}{|{\bf F_+}|^2|{\bf F_\times}|^2 - ({\bf F_+} \cdot {\bf F_\times})^2}\\
h^{(r)}_\times(t) = \frac{({\bf F_\times} \cdot {\bf \xi_\Delta(t)})(|{\bf F_+}|^2)-({\bf F_+} \cdot {\bf \xi_\Delta(t)})({\bf F_+} \cdot {\bf F_\times})}{|{\bf F_+}|^2|{\bf F_\times}|^2 - ({\bf F_+} \cdot {\bf F_\times})^2}
\end{matrix}
\right.
\end{equation}
%{\color{blue} WHY IS IMPORTANT TO SOLVE THE INVERSE PROBLEM?}
%{\color{blue} LEGGENDO LA PARTE SEGUENTE SEMBRA CHE IL PROBLEMA CHE STIAMO RISOLVENDO SIA STATO GIA' RISOLTO - SE SOSTITUISSIMO LA FRASE SOTTO SEMPLICEMENTE CON \emph{
Eq. \ref{Eq: solution} shows a degeneracy 
in the regions of the sky where the denominator of the two equations 
is near zero. 
% Here, we will not consider the case of the matrix deficiency.} }\emph{This solution has been already introduced in \cite{Rakhmanov}. As explained there, we expect to have a solution degeneracy 
%in the regions of the sky where the denominator of the two equations 
%is near zero.
In \cite{Rakhmanov} are proposed some solutions to avoid the matrix degeneracy, but we will not consider such approach
in the present study, since for the tested waveforms the statistic of cases affected by this deficiency is negligible, therefore it is not necessary to apply any regulator.\footnote{For 95\% of the sky the value of the denominator is greater than 0.01.} 
%we will
%use this approach in future studies.

To characterize the performance of the new algorithm, we use the Correlation Factor between the original waveform $h^{(i)}$ and 
the reconstructed  waveform $h^{(r)}$ for each polarization, defined as:
\begin{equation}\label{Eq. CorrFactor}
%\Delta_{+,\times}=\frac{\sum_t [h^{(r)}_{+,\times}(t) - h_{+,\times}(t)]^2}{\sum_t [h_{+,\times}(t)]^2}
C_{+,\times}=\frac{\left(h^{(r)}_{+,\times}, h^{(i)}_{+,\times}\right)}{\sqrt{\left(h^{(r)}_{+,\times},h^{(r)}_{+,\times}\right) \left(h^{(i)}_{+,\times},h^{(i)}_{+,\times}\right)}}
\end{equation}
where we define with $(,)$ the scalar product between two waveforms: $(a,b)=\int a(t)b(t) dt$. The Correlation Factor varies from -1 (opposite matching) to 1 (perfect matching).

The errors on polarization patterns are derived from the propagation of
errors of estimated quantities: 1) detector response; 2) sky direction; 3) polarization angle. %Note that the errors on sky direction propagate
It is fundamental to characterize the relative propagation of the errors  
to the polarization reconstruction, so we can understand which one of the just mentioned variables is predominant among the others.
We can disentangle the various contributions solving Eq. \ref{Eq: solution} in various
situations, i.e. 
\emph{by inserting iteratively the real parameters or the reconstructed ones.}  We consider three different cases: \textbf{a)} reconstructed detector response with true sky location and true polarization angle; 
\textbf{b)} reconstructed detector response and sky location with true polarization angle; 
\textbf{c)} reconstructed detector response, sky location and polarization angle.

%\begin{table} [h!]
%\begin{center}
%\begin{tabular}{|c|c|c|c|c|c|} 
%\hline  
%Case    &  Detector Response & Sky localisation  & Polarisation Angle & Colour\\
%\hline \hline 
%\textbf{a)} & reconstructed & injected & injected& \color{red}{\textbf{---}} \\
%\hline 
%\textbf{b)} & reconstructed & reconstructed & injected &  \color{green}{\textbf{---}}\\
%\hline 
%\textbf{c)} & reconstructed & reconstructed & reconstructed &  \color{yellow}{\textbf{---}}\\
%\hline \hline
%\textbf{d)} & reconstructed & reconstructed & reconstructed &  \color{blue}{\textbf{---}}\\
% & + delay & + delay&  + delay& \\
%%\textbf{d)} & \multicolumn{3}{c|}{all reconstructed + delay}\\
%\hline 
%\end{tabular}
%\caption{\small{Simple explanation of the situations considered in this work for disentangling the various contribution of 
%source errors. Each case uses reconstructed or injected quantities, according to the line in the table.} }
%\label{Tab. Cases}
%\end{center}
%\end{table}

\begin{figure}[!hbt]
\begin{center}
\includegraphics[width=80mm]{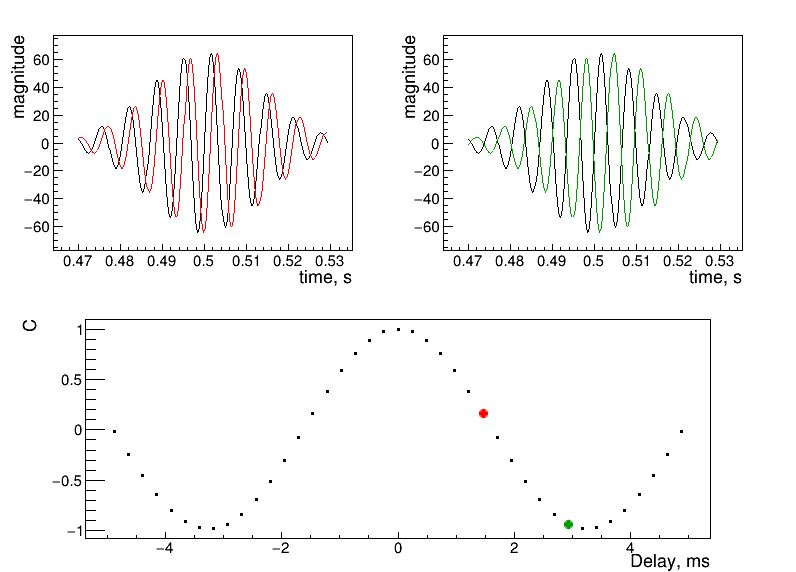}
\end{center}
\caption{\small{\textit{Time-shift effect on the correlation factor due to rigid shifts of a sinusoid with frequency of 153 Hz modulated 
by a Gaussian envelope. On top row, black curve is the original waveform, red/green (left/right) 
is the same waveform after a time-shift. On bottom row, values of $C$ calculated applying a time shift (x-axis) to
the original waveform. The red and green points refer to the examples in the top row.}}}
\label{Fig. Bias}
\end{figure}

\begin{table} [h!]
\begin{center}
\begin{tabular}{|c|c|c|c|} 
\hline  
\textbf{Case}    &  \textbf{ Detector } & \textbf{Sky}  & \textbf{  Polarisation  } \\
\textbf{  Color  }    &  \textbf{  Response  } & \textbf{  Localisation  }  & \textbf{Angle} \\
\hline \hline 
\color{red}{\textbf{a)}} & reconstructed & injected & injected \\
\hline 
\color{green}{\textbf{b)}} & reconstructed & reconstructed & injected \\
\hline 
\color{amber}{\textbf{c)}} & reconstructed & reconstructed & reconstructed \\
\hline \hline
\multirow{2}{*}{\color{blue}{\textbf{d)}}} & reconstructed & reconstructed & reconstructed \\
\cline{2-4}
% & + delay & + delay&  + delay \\
& \multicolumn{3}{c|}{+ time-shift}\\
\hline 
\end{tabular}
\caption{\small Sketch of the different cases presented in this paper to disentangle the various contribution of 
source errors. Each case uses reconstructed and/or injected quantities. }
\label{Tab. Cases}
\end{center}
\end{table}
We should remind that in this paper we are interested in any distortion between the original and reconstructed waveform, this is why we define the Correlation Factor, which is a suitable quantity for this problem. However, its value can decrease due to a possible time difference between the signal and the reconstructed waveform\\
%Despite the definition of Correlation Factor (Eq. \ref{Eq. CorrFactor}) be the proper variable to characterize distortions in the signal, we should keep in mind that the reconstruction is also affected by the time difference between the signal and the reconstructed waveform.  This key feature needs to be taken into account since the sky coordinate estimation has an effect also on the relative time-shift $\Delta_k$ among the involved detectors.
An example of this effect is reported in Fig. \ref{Fig. Bias}, where we have reported the values of $C$ when the reconstructed waveform is exactly the original one with different time-shifts. We can see that according to the applied time-shifts, 
the values of $C$ vary in the entire possible range of $[-1,1]$. This tells us that we should disentangle the distortions of the signal
with respect to any time-shifts in the calculation of the Correlation Factor.

We expect a similar behavior in the cases \textbf{b)} and \textbf{c)}.
In fact, case \textbf{b)} includes a time-shift (introduced by the different values of $\Delta_k$) between the original and reconstructed waveforms. In addition to case \textbf{b)}, case \textbf{c)} introduces the rotation of the waveform frame 
($h_+' = h_+ cos(\psi) + h_\times sin(\psi)$) which, especially for sinusoidal waveforms (see Fig. 1), is equivalent to a shift in time.
Hence, in case c) we have two time-shifts to take into account, one that is the same of case b) and the other due to the difference between original and reconstruted polarization angle
The interesting challenge is: do we have a distortion in the signal when we estimate the wrong sky position? Or is an effect due to a rigid shift as
in Fig. \ref{Fig. Bias}? To answer the question, we include the case \textbf{d)}, 
in which we reconstruct the detector response, the sky localization and also the polarization angle (as in case \textbf{c)} ) and then we add a time shift on $h_+$ and $h_{\times}$.
The applied time shift is the maximum value of the cross-correlation function of time
 in which we calculate the cross-correlation between the
injected and reconstructed polarization patterns. Hence, we estimate the possible shift among the two waveforms and
we calculate the $C$ value after correcting the reconstructed polarizations with this time-shift. 
In this way we can focus only on possible signal distortions.

\section{Monte Carlo simulations}\label{Sec: sim}
As starting point for our tool we decide to take the results obtained by the GW transient signal algorithm in use by  the LIGO and Virgo collaborations called coherent Waveburst \cite{cWB2008, cWB2015}. 
It is an algorithm to measure energy excesses  over the detector noise in the time-frequency domain and 
combining these excesses coherently among the various detectors. This is performed introducing a maximum likelihood
approach to define the ratio among the probability of having a signal in the data over the probability of only noise.
This does not need a particular
assumption on the expected waveform, making it open to a wide class of transient signals.

The algorithm has
recently improved in preparation for the Advanced Detector Era. 
%{\color{blue} QUI NON TI SEGUO, DALLA FRASE SOTTO HO L'IMPRESSIONE CHE QUESTA ANALISI SIA PARTE DI CWB, COME SI CHIEDEVA IL REVISORE}
The main improvement concerns a new method for the estimation of the event parameters which considers assumptions on
the polarization state (circular, linear, elliptical, etc...) \cite{cWB2015}. This improvement is 
particularly suited for this work, because, despite it does not calculate directly the two plus and cross polarizations, it is
implicitly connected to them through the calculation of the polarization state.
The algorithm performances on sky localization and detector response are 
reported in the previous results \cite{PRD2011, Reed}.

For this work we contemplate a network composed of three interferometers: the two LIGO (L1, Livingston and H1, Hanford) ones 
and the Virgo (Cascina, Italy) detector, with simulated Gaussian detector noise considering the amplitude spectral density at the design sensitivity 
\cite{LIGO psd, Virgo psd}. 
Even though most of the noise background in detectors is Gaussian, random instrumental artifacts can make 
the background far from Gaussian \cite{S5,S6,O1}. The use of Gaussian noise is just the starting point to verify the performances of this new tool. 
We expect that glitches would affect only the detection confidence and not the reconstruction of the waveforms, as explained in the case of the sky localization \cite{Reed}.
We will check the real behavior of noise when we will handle the data analysis of the three Advanced detectors.

%{ \color{blue} Linear combinations of detector outputs with no gravitational signal eliminate almost all of them because the glitches are not correlated in the data streams of separated detectors. 
% --LA FRASE SOTTO LA CAMBIEREI COSI' }
%{\color{red} NON SONO D'ACCORDO CON SCRIVERE QUELLA PARTE LI' IN BLU. (la parte precedente mi va bene, e ho tolto il blu)
% E' vera ma non e' focalizzata sul problema, perche' comunque rimangono eventi
%non gaussiani, anche pur facendo quello che e' scritto li'. E' piu' il fatto che comunque anche se rimangono i glitches non e' che danno
%direttamente fastidio agli eventi stessi, o comunque in minima parte. E' quello che volevo dire con la frase sotto:}
%{\color{red} \emph{this has
%a major effect on the detection. However, for signal limited in time and frequency, it is really improbable that a glitch interfers with the signal.}} {\color{blue} SCUSA MA NON CAPISCO CHE COSA VUOI METTERE DI PIU' SPECIFICO, MI PARE CHE QUANTO ESPRIMI IN ROSSO SIA CONTENUTO NEL TESTO BLU, MA FORSE NON HO CAPITO CHE INTENDI}
We injected the so-called sine-Gaussian and WhiteNoiseBurst waveforms, which are among the standard tested waveforms for burst searches, respectively representing the limiting case for most simple and complicated morphologies. The former are defined
as follows:
\begin{equation}
\begin{split}
h_+(t) &= h \frac{1+\cos^2(\iota)}{2} sin(2\pi t f_0) exp(-t^2/\tau^2)\\
h_\times(t) &= h \cos(\iota) cos(2\pi t f_0) exp(-t^2/\tau^2)
\end{split}\label{Eq. SG}
\end{equation}
where $f_0$ is the central frequency, $\tau$ is related to the waveform quality factor $Q=\sqrt{2}\pi f_0 \tau$ and
the inclination angle $\iota$ is uniformly distributed. For this study
we took into account three quality factors $Q=3,9,100$ and three central frequencies: $f_0=235, 554, 1053$ Hz, 
with source coordinates uniformly distributed in the sky.
White Noise Bursts are frequency band limited white noise with a time Gaussian envelope. They have no particular polarization,
while SineGaussians have elliptical polarization.
To distinguish  the various sources of error, we compute  the correlation factor (Eq. \ref{Eq. CorrFactor}) 
for the different cases 
\textbf{a)}, \textbf{b)}, \textbf{c)} and \textbf{d)}.
To characterize the algorithm performances, we inject uniformly in the sky at discrete values of the
network SNR: 10, 12, 15, 20, 25, 30, 35, where network signal-to-noise ratio (SNR) is defined as the square sum of the ratio of the reconstructed waveform in the frequency domain $(\tilde{h}_+,\tilde{h}_\times)$ and the amplitude spectral density $S_k(f)$ of each detector $k$:
 $SNR^2= \sum_k \int \frac{\tilde{h}_+^2+\tilde{h}_\times^2}{S_k(f)} df$.

\section{Results}\label{Sec: results}

\begin{figure*}[!hbt]
\begin{center}
\begin{tabular}{cc}
\includegraphics[width=80mm]{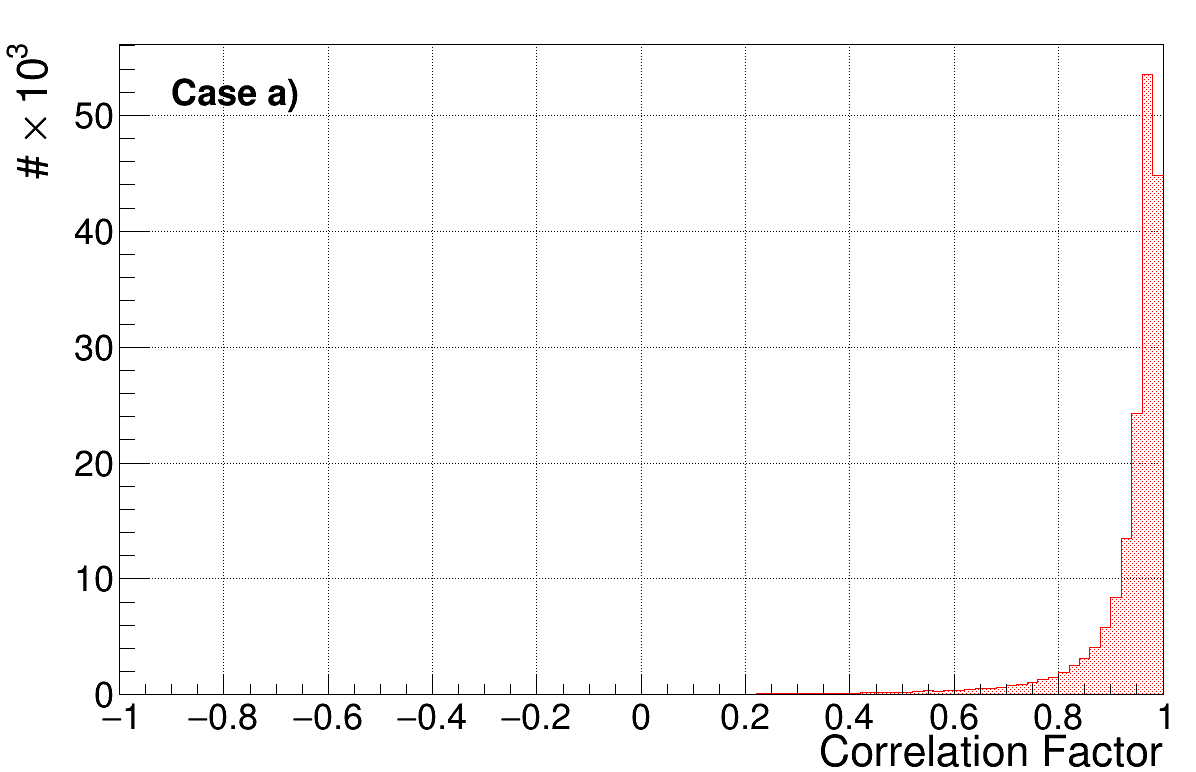} &
\includegraphics[width=80mm]{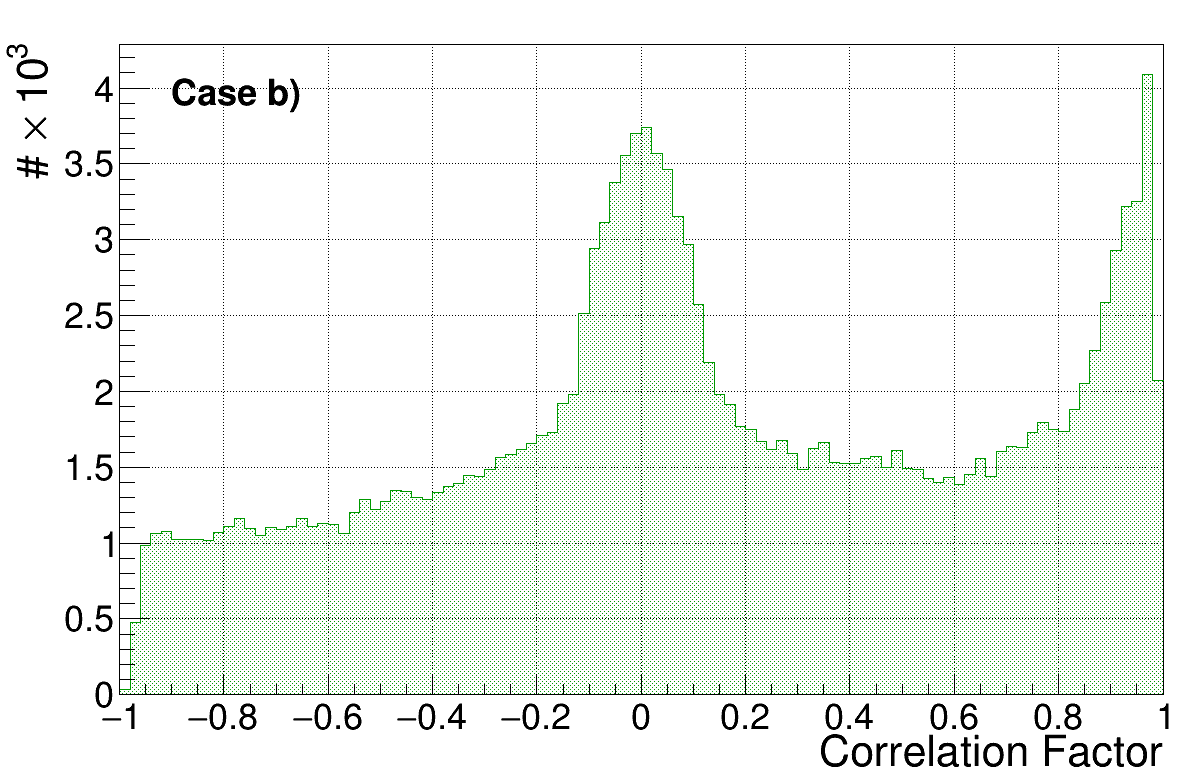} \\
\includegraphics[width=80mm]{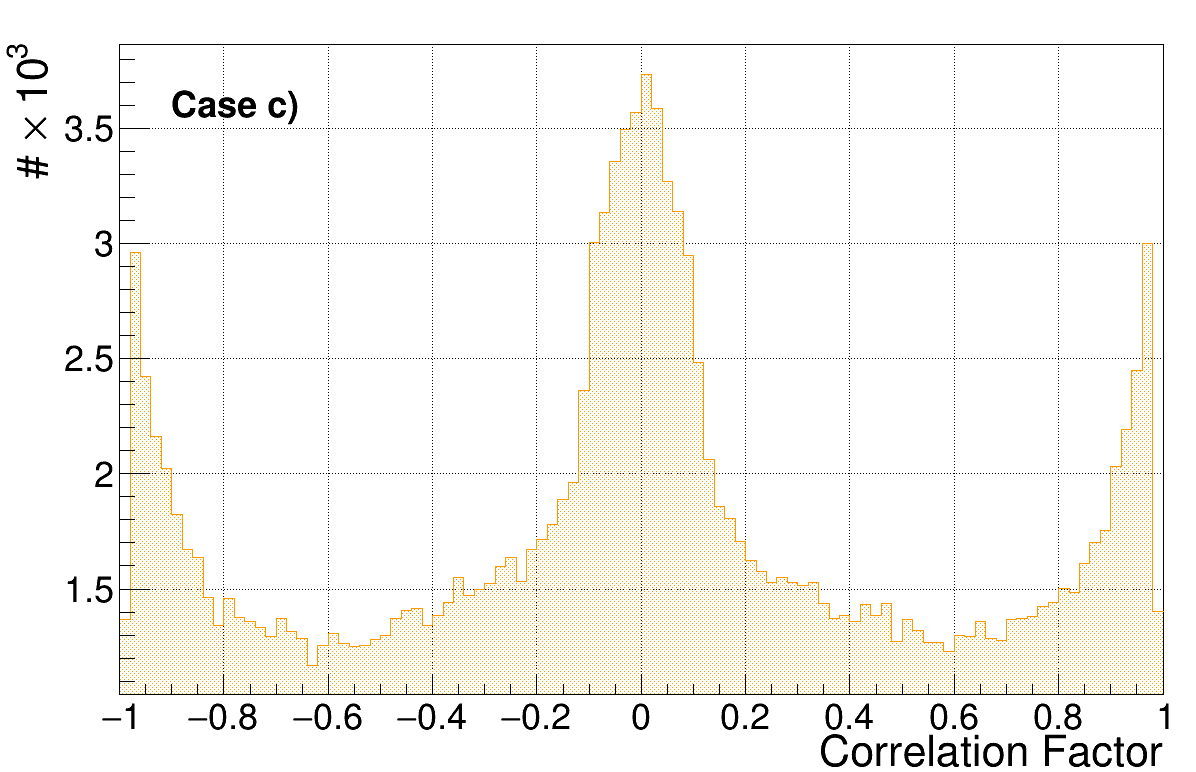} &
\includegraphics[width=80mm]{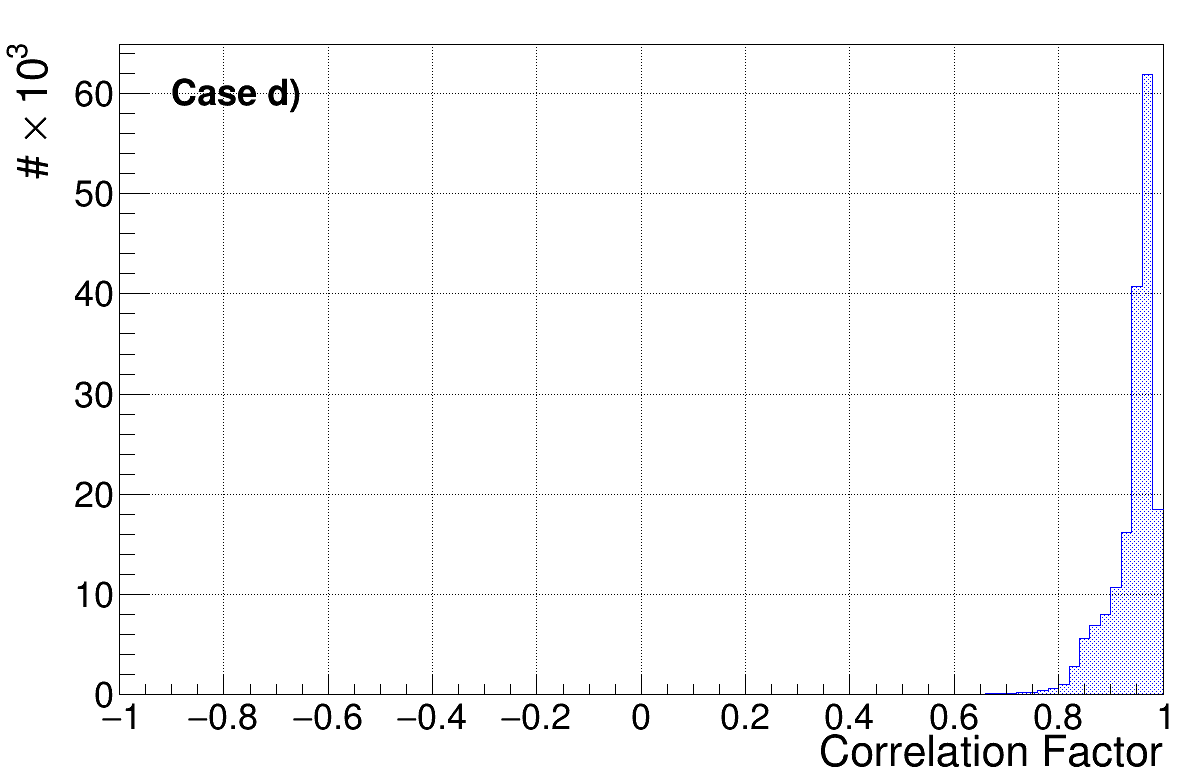}
\end{tabular}
\end{center}
\caption{\small{\textit{Correlation factor of the plus polarization for injections of a SineGaussian with 
central frequency of 253 Hz and Q=9 uniformly distributed in the sky and different values of SNR. 
From left to right, top to bottom: case \textbf{a)}, \textbf{b)}, \textbf{c)}, \textbf{d)}.}}}
\label{Fig. Caseresult}
\end{figure*}
\begin{figure*}[!hbt]
\begin{center}
\begin{tabular}{ccc}
\includegraphics[width=55mm]{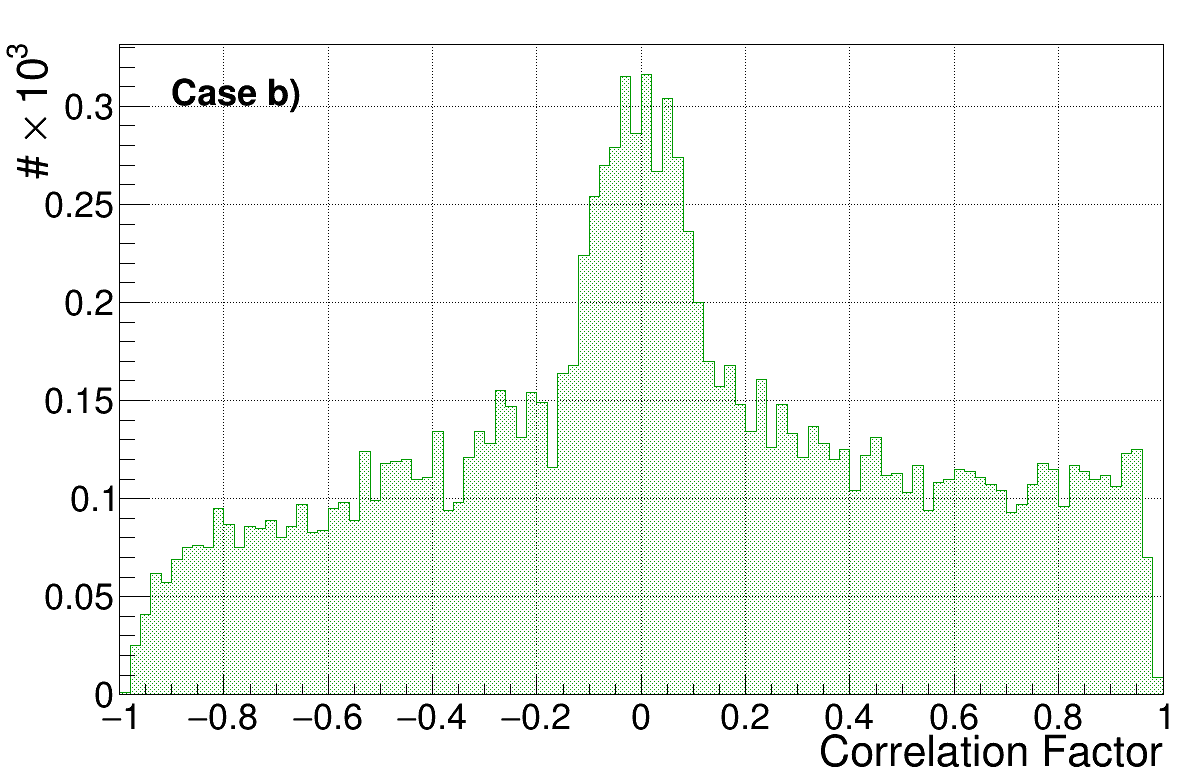} &
\includegraphics[width=55mm]{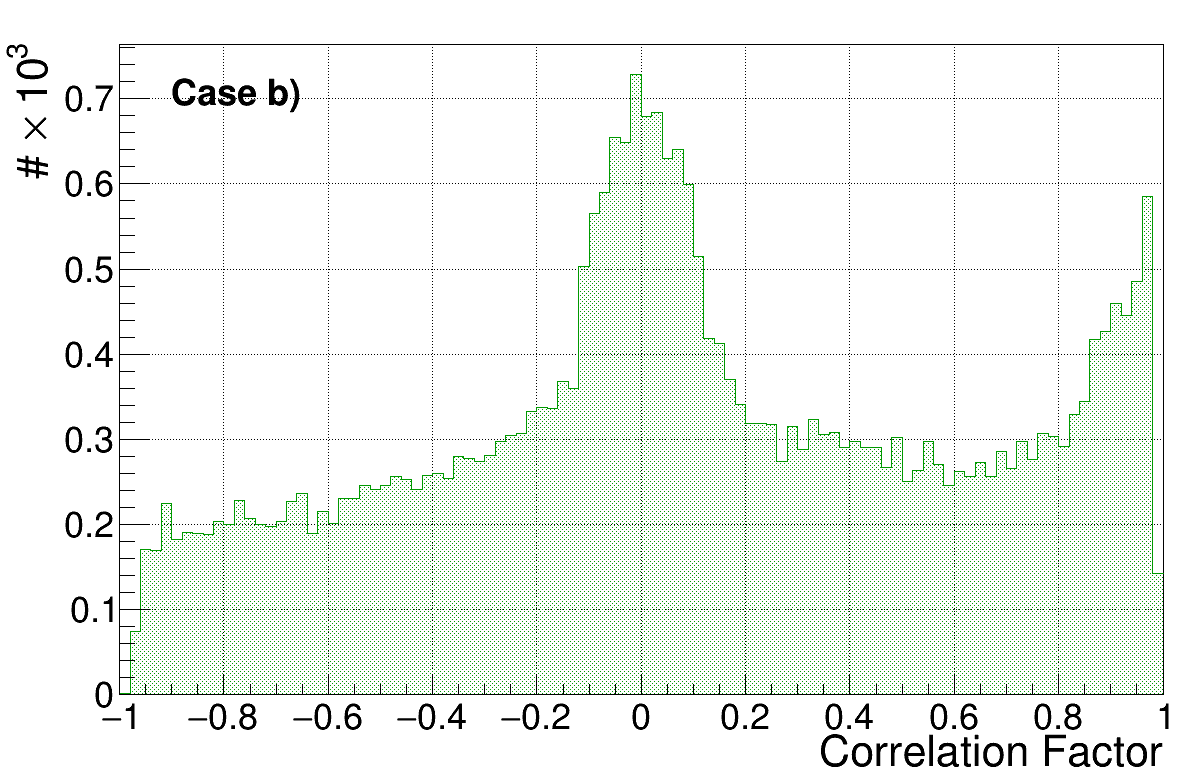} &
\includegraphics[width=55mm]{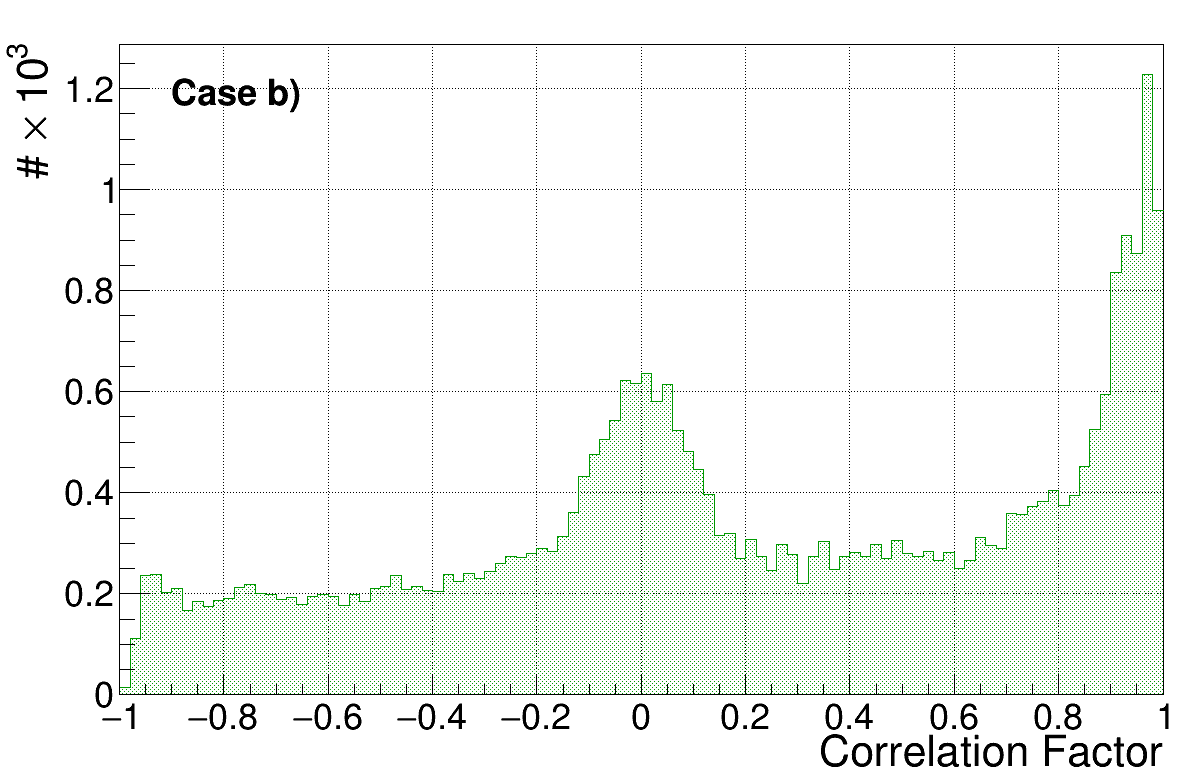}
\end{tabular}
\end{center}
\caption{\small{\textit{Correlation factor of the plus polarization for case b)
 for injections of a SineGaussian 
with central frequency of 253 Hz and Q=9 uniformly distributed in the sky and selected
values of SNR. From left to right: SNR=12, 20, 35.}}}
\label{Fig. DiffSNR}
\end{figure*}
\begin{figure*}[!hbt]
\begin{center}
\begin{tabular}{ccc}
\includegraphics[width=55mm]{SGE235Q9_hp_allfactors_angle.png} &
\includegraphics[width=55mm]{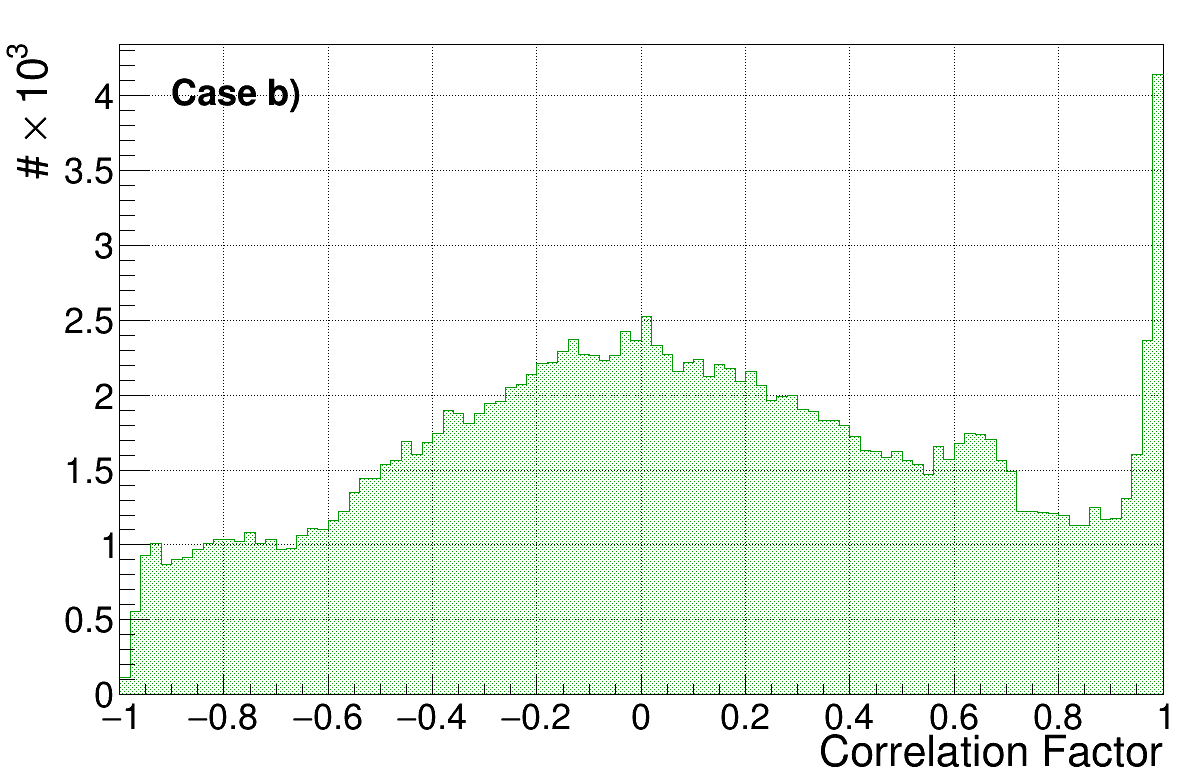} &
\includegraphics[width=55mm]{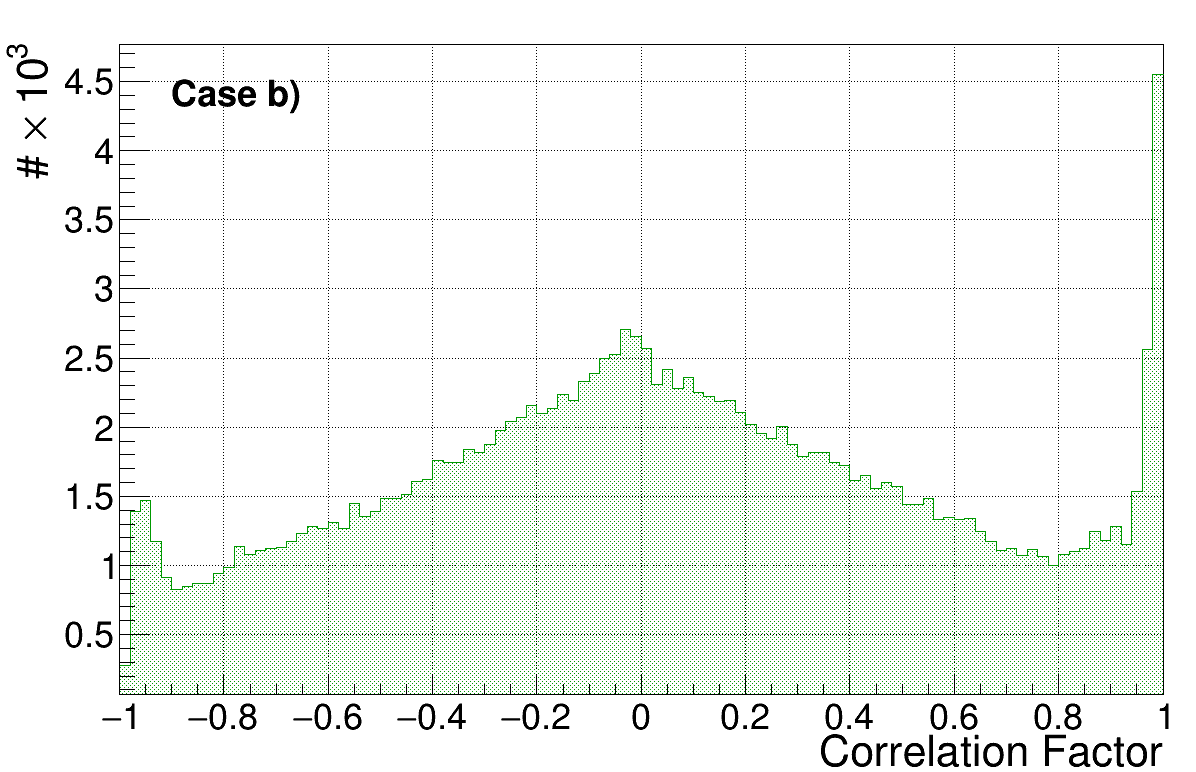}
\end{tabular}
\end{center}
\caption{\small{\textit{Correlation factor of the plus polarization for case b)
 for injections of a SineGaussian
with Q=9 uniformly distributed in the sky and selected
central frequencies. From left to right: central frequency 235, 554, 1053 Hz.}}}
\label{Fig. DiffFreq}
\end{figure*}

The aim of the new presented tool is the reconstruction of both polarizations ($+$,$\times$) starting from the results 
(detector responses, sky localization) of a standard GW transient algorithm.
We started the study disentangling the contribution of detector response and sky localization on the estimation of the polarizations.
%{\color{blue} WHY DO WE HAVE THIS HYPOTHESIS? }
%{\color{blue} QUESTA L'HO RIMESSA NON SO SE L'AVEVI TOLTA PERCHE' AVEVI CAPITO MALE O PERCHE' NON TI PIACEVA -We would expect the polarization reconstruction to be more sensitive to sky localization errors than to  single detector response errors. This can be verified looking at the difference of the results between the case \textbf{a)}  and \textbf{b)-c)}. }
In Fig. \ref{Fig. Caseresult} we report an example of the $C_+$ distribution 
(Eq. \ref{Eq. CorrFactor}) for an elliptical SineGaussian centered at 235 Hz and Q=9 for all the injected SNR. 
In case \textbf{a)}, where we have reconstructed only the detector response, 
the distribution of the correlation factor is near 1, confirming that the detector response estimation
is in agreement with the expected waveforms for the complete set of SNRs. 
%{\color{red} WHY DO WE HAVE NEW POPULATIONS??? 
In case b), where we reconstruct both the detector response and the sky positions, it
 appears a new population
centered in 0. %{\color{red}This behavious can be due to two different effects: signal distortion, time-shift.}
Whereas, for case  c), where in addition to the information in case b)
we reconstruct also the polarization angle,
%{\color{red} where we are reconstructing also the polarization angle}, 
there are three main populations centered respectively to -1, 0, 1.
The appearance of these new peaks in b) and c) can be due to two different effects: 
signal distortion and/or time-shift. As we have explained in Fig. \ref{Fig. Bias}, introducing time-shifts lower the value of the Correlation Factor.
When we look at case \textbf{d)}, where we avoid a possible time shift bringing back the c) waveform to the right time, 
results display only a remaining distribution around 1, similar to what happens in case a). 
This answer the question we were posing in the previous section: does a wrong sky localization create distortion of the original signal? 
Results show that the main effect of estimating the wrong sky position is the introduction of a time-shift between the estimated polarization and the original one, but such distortions on the waveform shape are negligible.
The fact that for case b) we do not have so many values of Correlation Factor less than 0 means that the errors
from the sky localization cause a time-shift to the reconstructed waveform that is not bigger than a certain value. Thinking
in term of phase shift, it is not bigger than a 90 degrees shift.\footnote{For sinusoidal waveforms, a generic time-shift is equivalent to a phase shift.}
To confirm this last sentence, we compare the results related to case b) 
considering different injection SNRs (Fig. \ref{Fig. DiffSNR} ) and SineGaussian with 
different frequencies and same SNR (Fig. \ref{Fig. DiffFreq}). 

Fig.  \ref{Fig. DiffSNR} , where we reported 3 different cases,
SNR=12, SNR=20, SNR=35, displays that the widths of the peaks are not different in the three cases. This verifies that the peak at 0 is not
related to the SNR, but it gives an hint that could be related to the sample rate.
However, the fact that the height of the peaks in 0 and in 1 changes for different SNRs tell us that the sky reconstruction
is better for high SNR (as expected).
More the SNR is high, less is the number of events affected by a wrong sky localization,
which produces a time-shift between the injected and reconstructed waveform
polarizations resulting in low values of the Correlation Factor.
\\
In Fig \ref{Fig. DiffFreq} we show the performances for waveforms at 554 Hz and 1053 Hz of 
central frequency and compared these with the one of 253 Hz already reported in Fig \ref{Fig. Caseresult}.
We can see that the distribution of the results are slightly different, in particular, when the
frequency increases, the distribution is not anymore showing the peak in 0, but 
the values of the correlation factor are more spread in the possible range [-1,1]. The reason
is that for
this work we adopted the standard SineGaussian definition \cite{S5, S6}, 
which groups in the same set waveforms with same quality factor $Q=\sqrt{2}\pi f_0 \tau$ 
(see Eq. \ref{Eq. SG}). The quantity Q is mainly related to the number of cycles
that characterize the SineGaussian. Therefore, in the same set, given that
we have equal cycles but different frequencies, the time duration of the waveforms is
inversely proportional to the frequency.
%, as we see in the Fig. \ref{Fig.Q9}, where we
%have reported three different waveforms with identical Q=9.
Increasing the number of cycles means the possibility to explore more phase shifts between 
the injected and the reconstructed waveform, when we apply a generic time-shift between the 
two waveforms. The time-shift between injected and reconstructed depends only on the
sky location and, even if the reconstructed sky localization is better when the frequency increases
\cite{PRD2011, Fairhust}, such improvement is not enough to contrast the effect of having different 
number of cycles coming from waveforms at higher frequencies. 
Increasing the number of cycles means the possibility to explore more phase shifts between 
the injected and the reconstructed waveform,
that shapes the distribution of values of the Correlation Factor.

\begin{figure*}[!hbt]
\begin{center}
\begin{tabular}{ccc}
\includegraphics[width=55mm]{SGE235Q9_hp_allfactors_psi.png} &
\includegraphics[width=55mm]{SGE235Q9_hp_allfactors_psi.png} &
\includegraphics[width=55mm]{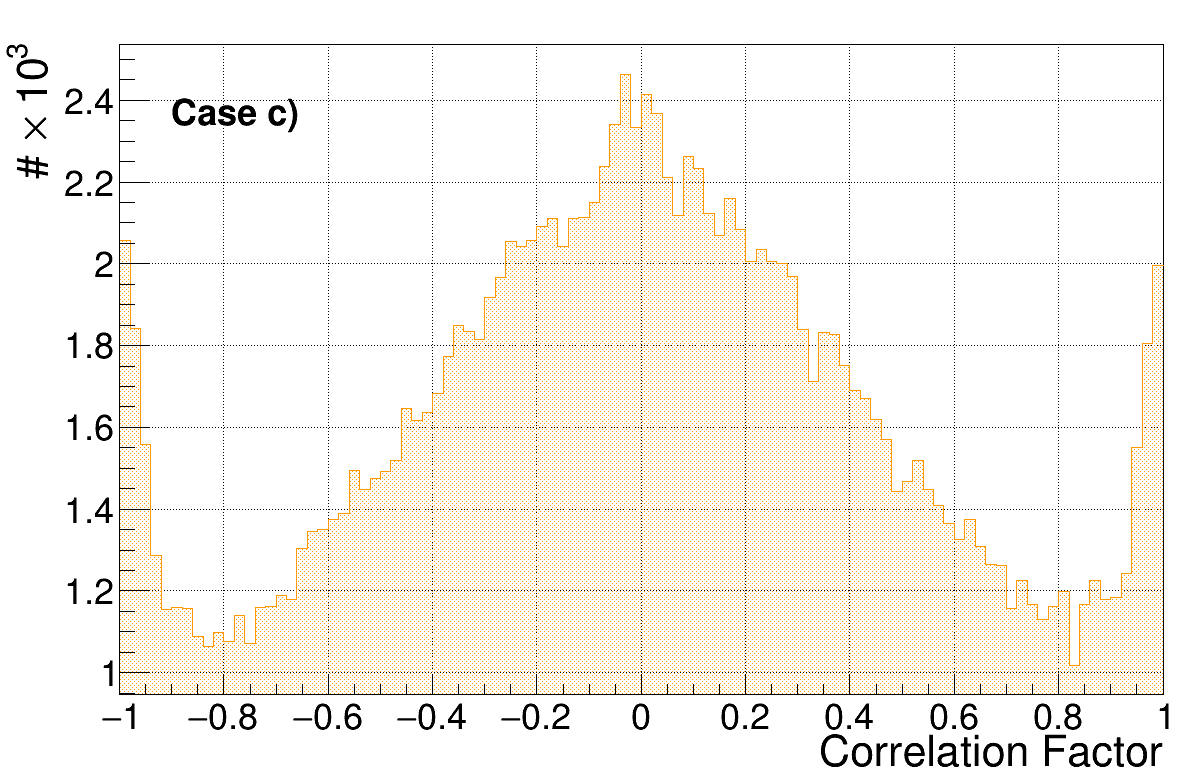}
\end{tabular}
\end{center}
\caption{\small{\textit{Correlation factor of the plus polarization for case c)
 for injections of a SineGaussian
with Q=9 uniformly distributed in the sky and selected
central frequencies. From left to right: central frequency 235, 554, 1053 Hz.}}}
\label{Fig. DiffSNR_case_c}
\end{figure*}

\begin{figure}[!hbt]
\begin{center}
\includegraphics[width=80mm]{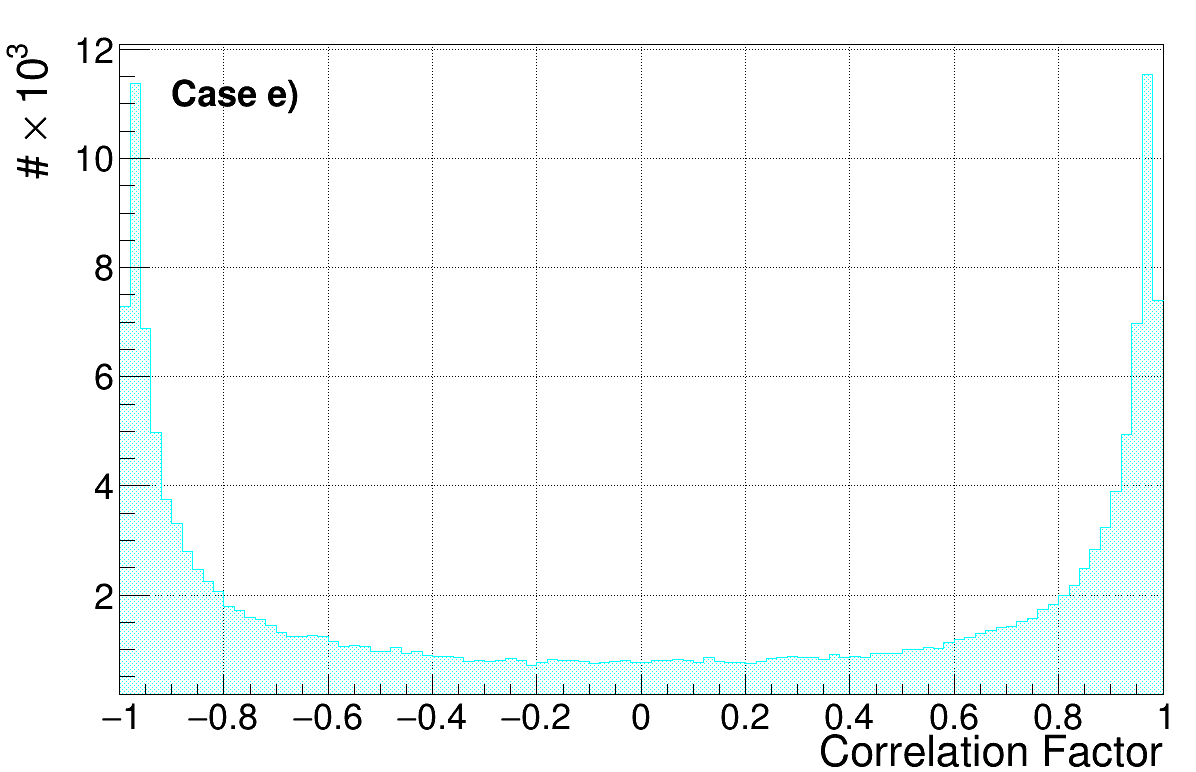}
\end{center}
\caption{\small{\textit{Correlation factor of the plus polarization for case c) 
for injections of a SineGaussian
with central frequency of 253 Hz and Q=9 uniformly distributed in the sky and all the 
tested SNR .}}}
\label{Fig. case_e}
\end{figure}

For case c), instead, the reconstructed polarization angle can differ from the original one in the complete range between 0 and 360 degrees,
for this reason it appears the peak at -1. The accumulation point in -1 is always present, 
for each SNR and frequency, as we can see in the Fig. \ref{Fig. DiffSNR_case_c}. 
To confirm this we tried the case of injected sky position and reconstructed polarization angle
(we call this case e) and we report the result for the SineGaussian at central frequency
of 235 Hz and Q=9 in Fig \ref{Fig. case_e} (but for other waveforms results are similar).
The Fig. \ref{Fig. case_e} shows that the peak at -1 is still
present, validating the fact that it is related to the wrong estimation of the polarization angle (moreover
the peak at 0 is disappeared).

\begin{figure}[!hbt]
\begin{center}
\subfigure[SineGaussian with central frequency of 253 Hz and Q=9\label{Fig. Caseresult 0d5 SGQ9}]{\includegraphics[width=88mm]{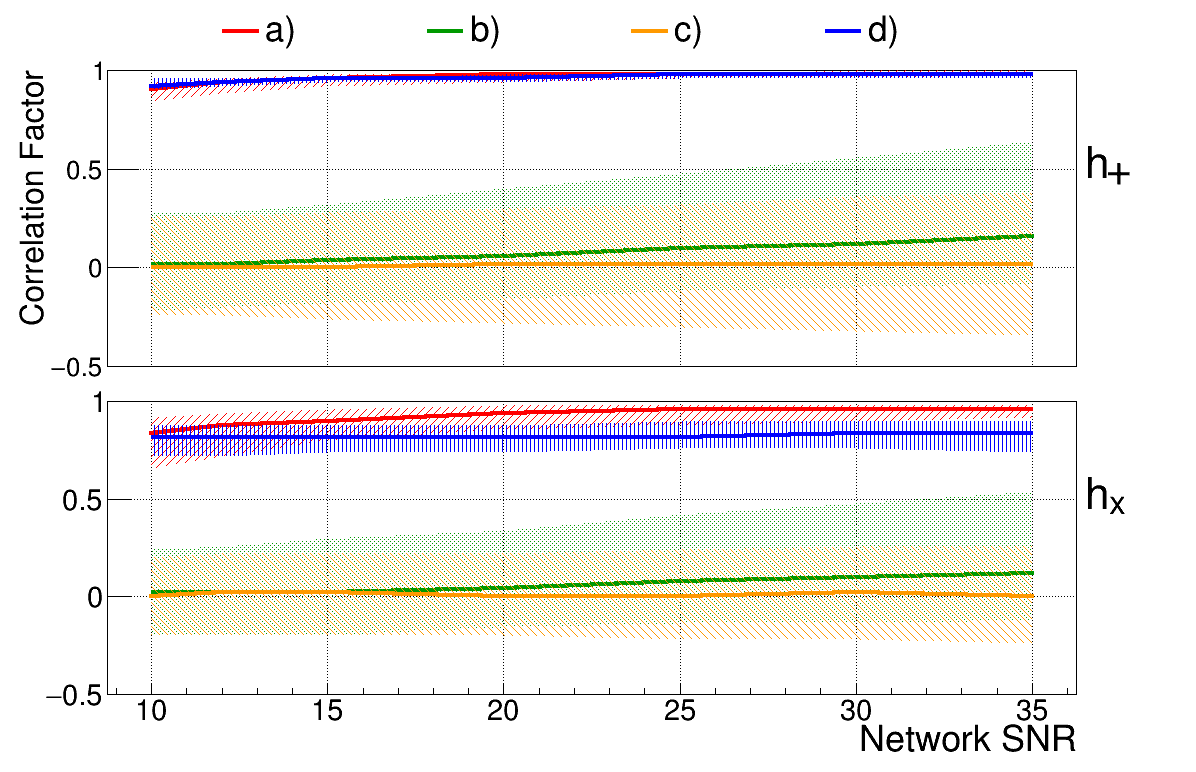}}\\
\subfigure[WhiteNoiseBurst with frequency range (250,350) Hz  and duration of 100ms.\label{Fig. Caseresult 0d5 WNB}]{\includegraphics[width=88mm]{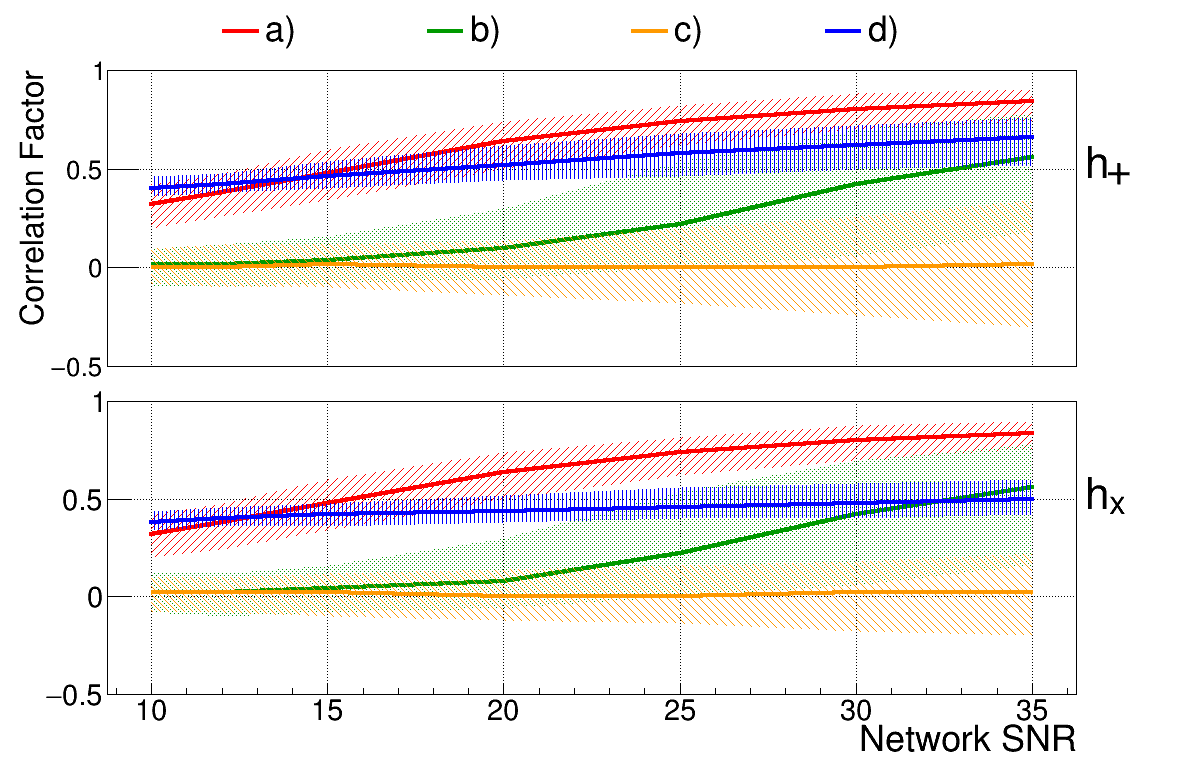}}\\
\subfigure[Binary black holes coalescence with single component masses between 15 and 25 solar masses\label{Fig. Caserresult 0d5 CBC}]{\includegraphics[width=88mm]{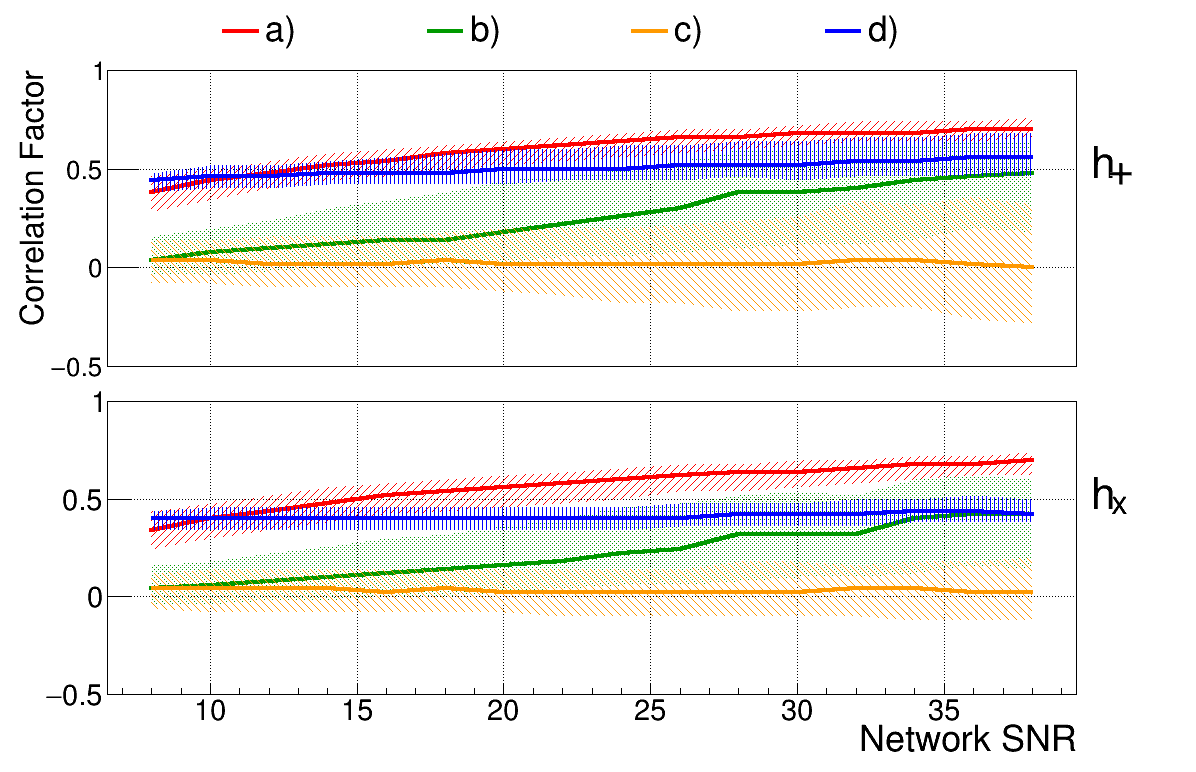}}
\end{center}
\caption{\small{\textit{Median (lines) and 30\%-70\% percentile (colored regions) of Correlation Factor for the plus (top) and cross (bottom) 
polarization for injections  uniformly distributed in the sky as a function of the injected SNR.
Colours refer to cases labelled in the text as \textbf{a)} (red), \textbf{b)} (green), \textbf{c)} (orange), \textbf{d)} (blue).}}}
\end{figure}

On Fig. \ref{Fig. Caseresult 0d5 SGQ9} we estimate, from the distribution of Fig. \ref{Fig. Caseresult},
the median value. The shaded regions refer to the values for which we obtain 
%the values of 
the percentile between 30\% and 70\%. 
On \ref{Fig. Caseresult 0d5 WNB} we have the same results for a WhiteNoiseBurst. 
We see that for each case (from \textbf{a) to d)}) a waveform like a SineGaussian shows in general a median
nearer to the optimal value of 1 than a waveform like a WhiteNoiseBurst. 
These waveforms show a time-frequency representation which involves a greater number of pixels. Having more pixels
naturally increases the noise contribution to the waveform reconstruction. This easily explains the reason why performances
for SineGaussian are better than for WhiteNoiseBurst, given that it is possible to characterize the first ones with less 
time-frequency pixels. Indeed, for the SineGaussian case, results are independent of the tested SNR. Since a decrease in the SNR values produces a contamination of the noise in the performances, we chose a network SNR=10 (detector SNR$\approx10/\sqrt{3}<6$) as a reasonable threshold for a candidate event, as explained and done before for searches in \cite{burstPaper}.

%We report the localization error as the angular difference between estimated and real sky direction of the source, and the 
%error on detector response.

%\begin{figure}[!hbt]
%\begin{center}
%%\begin{tabular}{cc}
%%\includegraphics[width=42mm]{SGE235Q9_hp_0d5.png} &
%%\includegraphics[width=42mm]{SGE235Q9_hc_0d5.png}\\
%%\includegraphics[width=42mm]{WNB250_100_0d100_hp_0d5.png} &
%%\includegraphics[width=42mm]{WNB250_100_0d100_hc_0d5.png}\\
%%\end{tabular}
%\includegraphics[width=90mm]{SGE235Q9_hphc_0d5.png}\\
%\includegraphics[width=90mm]{WNB250_100_0d100_hphc_0d5.png}
%\end{center}
%\caption{\small{\textit{CF50\% of the plus (left) and cross (right) polarization for injections  uniformly distributed 
%in the sky as a function of the injected SNR.
%On the top a SineGaussian with central frequency of 253 Hz and Q=9 while on the bottom a WhiteNoiseBurst with frequency 
%range [250,350] Hz  and duration of 100ms.
%Coloured lines refer to cases labelled in the text as \textbf{a)} (red), \textbf{b)} (green), \textbf{c)} (orange), \textbf{d)} (blue).}}}
%\label{Fig. Caseresult 0d5}
%\end{figure}

\begin{figure}[!hbt]
\begin{center}
\subfigure[SineGaussian with Q=9 and central frequency at 235 Hz (black), 554 Hz (red), 1053 Hz (green).\label{Fig. SGQ9 0d5}]{\includegraphics[width=85mm]{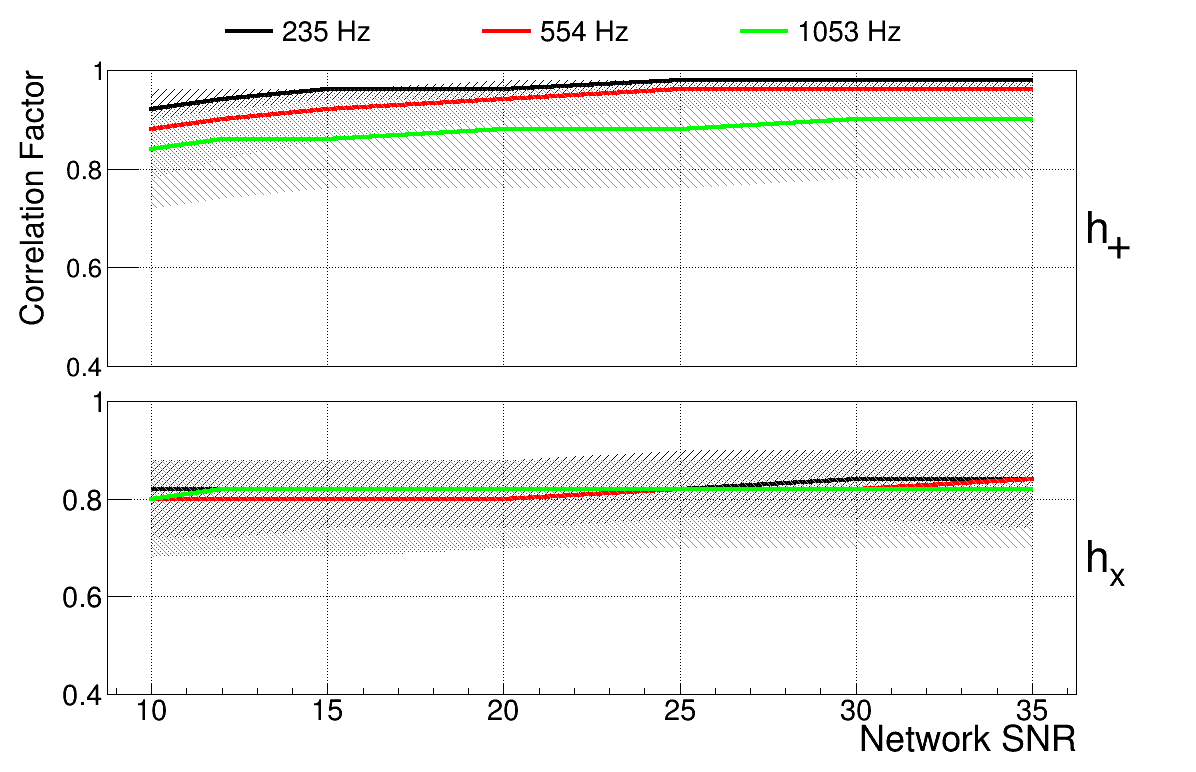}}\\
\subfigure[SineGaussian with central frequency = 235 Hz and Q=3 (black), Q=9 (red) , Q=100 (green).\label{Fig. 235 0d5}]{\includegraphics[width=85mm]{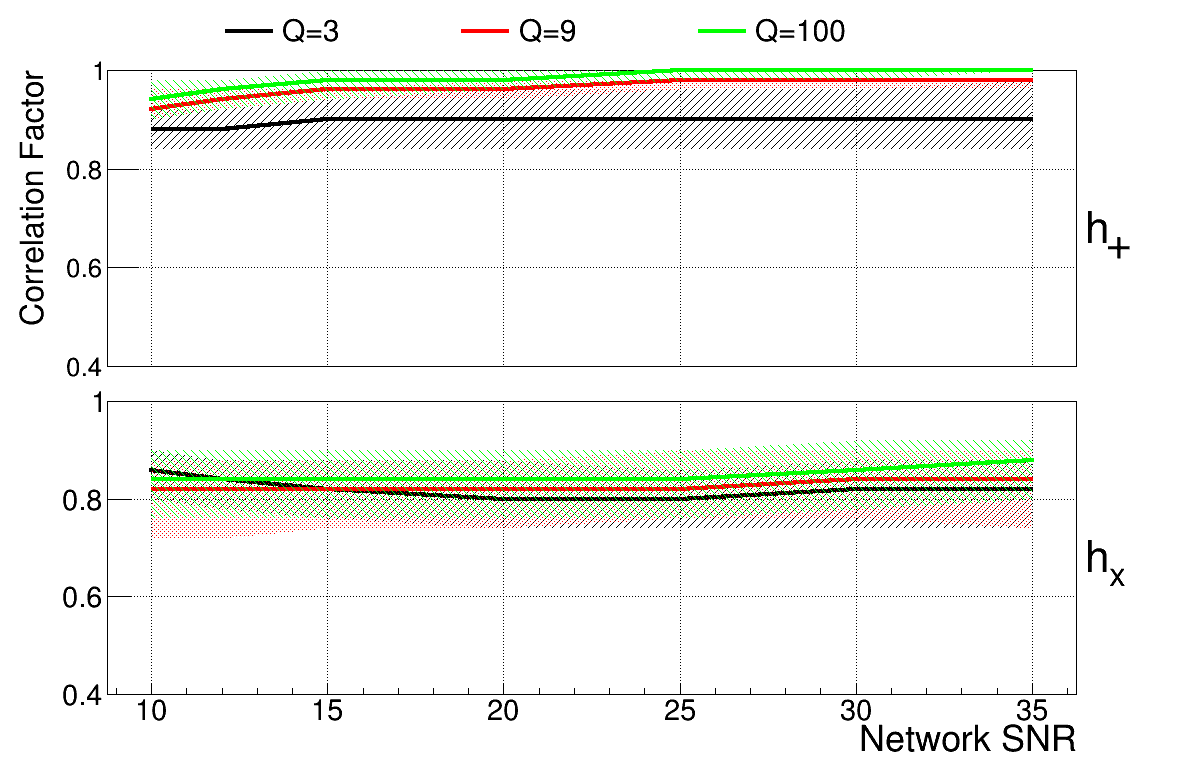}}\\
\subfigure[WhiteNoiseBurst with bandwidth (250,350) Hz (black), (500,600) Hz (red) and (1000,1100) Hz (green).\label{Fig. WNB 0d5}]{\includegraphics[width=85mm]{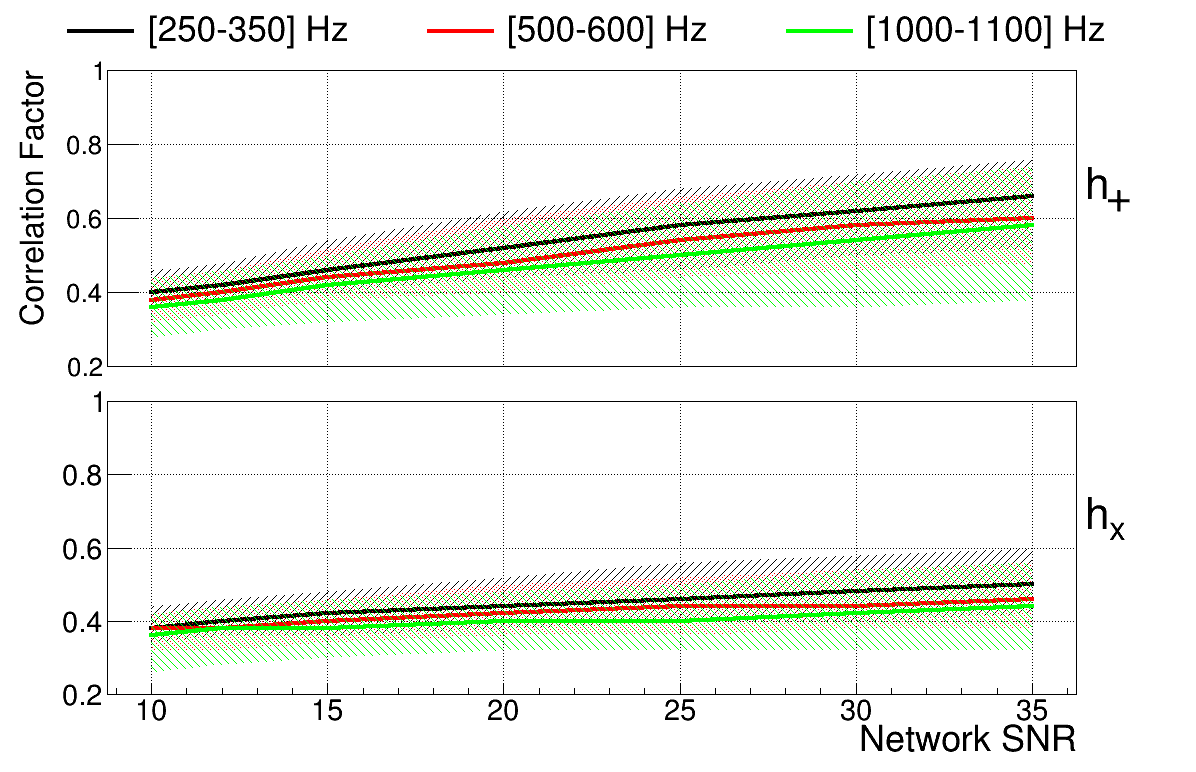}}
\end{center}
\caption{\small{\textit{Median (lines) and 30\%-70\% percentile (colored regions) of Correlation Factor for the plus (top) and cross (bottom) 
polarization for different signals for the \textbf{d)} case.
The y-axis shows what is the correlation factor value for which the 50\% (lines) and the 30\%-70\% percentiles of the recovered waveforms 
have bigger cross correlation.}}}
\end{figure}

\subsection{Transient signals}

In this paper we are interested in the reconstruction of the waveform shape with the aim to estimate the characteristics of the generating source from the waveform itself. 
Given the purpose, we want to focus on eventual signal distortions in the reconstructed polarization patterns. 
Since in real analysis we do not know the true quantities, we should focus on case  \textbf{c)}.
However, as already discussed and shown in Fig. \ref{Fig. Caseresult} for this case the correlation factor $C$ describes both the effects of signal distortion
and time-shift. Hence, in the following we consider the case \textbf{d)} because it is connected only to the distortion and not to any eventual time-shift applied to the signal.\\
In Fig. \ref{Fig. SGQ9 0d5} we report the comparison of SineGaussian with the same Q=9 but at the different central frequencies. We can see that 
performances are better for plus polarization, $(+)$. This is probably because for most of the injection the $(+)$ plus polarization 
has higher energy than the cross one, $(\times)$. Indeed there are negligible differences for the $(\times)$ cross polarization among the 
three frequencies, but for the plus polarization $(+)$ we can see that performances are better for lower frequencies, probably due 
to the fact that these oscillations are less separated in time.
%\begin{figure}[!hbt]
%\begin{center}
%%\begin{tabular}{cc}
%%\includegraphics[width=42mm]{Q9_hp_delay_0d5.png} &
%%\includegraphics[width=42mm]{Q9_hc_delay_0d5.png} 
%%\end{tabular}
%\includegraphics[width=90mm]{Q9_hphc_delay_0d5.png}
%\end{center}
%\caption{\small{\textit{CF50\% of the plus (left) and cross (right) polarization for SineGaussian injections 
%with Q=9 at different frequencies for the \textbf{d)} case.
%The y-axis shows what is the correlation factor value for which the 50\% of the recovered waveforms have bigger cross correlation.
%Coloured lines refer to SineGaussian with central frequency at 235 (black), 554 (red), 1053 (green) Hz.}}}
%\label{Fig. SGQ9 0d5}
%\end{figure}
This is confirmed also in Fig. \ref{Fig. 235 0d5} where we show the comparison for SineGaussian with same central
frequency but different Q. We see that when Q increases, performances are better, because waveform's bandwidth is narrow 
and the time-frequency representation involves less pixels.

%\begin{figure}[!hbt]
%\begin{center}
%%\begin{tabular}{cc}
%%\includegraphics[width=42mm]{235_hp_delay_0d5.png} &
%%\includegraphics[width=42mm]{235_hc_delay_0d5.png}
%%\end{tabular}
%\includegraphics[width=90mm]{235_hphc_delay_0d5.png}
%\end{center}
%\caption{\small{\textit{CF50\% of the plus (left) and cross (right) polarization for SineGaussian injections 
%with central frequency at 235 Hz and different Q for the \textbf{d)} case.
%The y-axis shows what is the correlation factor value for which the 50\% of the recovered waveforms have bigger cross correlation.
%Coloured lines refer to SineGaussian with Q=3 (black), 9 (red) , 100 (green).}}}
%\label{Fig. 235 0d5}
%\end{figure}

Similar results for WhiteNoiseBurst are shown in Fig. \ref{Fig. WNB 0d5}, where the performances are better
for lower frequencies. Moreover, for these waveforms the performances are slightly worse, 
with respect to SineGaussian. Indeed, the WhiteNoiseBurst signals are more widespread in the time-frequency domain than
the Sine-Gaussian, which makes more difficult to accurately reconstruct the complete waveform.
 As already shown (Fig. \ref{Fig. Caseresult 0d5 WNB}), for these waveforms the contribution to the errors coming from the
detector response estimation is more important than the SineGaussian. 

%\begin{figure}[!hbt]
%\begin{center}
%%\begin{tabular}{cc}
%%\includegraphics[width=42mm]{WNB_hp_delay_0d5.png} &
%%\includegraphics[width=42mm]{WNB_hc_delay_0d5.png}
%%\end{tabular}
%\includegraphics[width=90mm]{WNB_hphc_delay_0d5.png}
%\end{center}
%\caption{\small{\textit{CF50\% of the plus (left) and cross (right) polarization for WhiteNoiseBurst injections 
%with different bandwidth and duration of 100 ms for the \textbf{d)} case.
%The y-axis shows what is the correlation factor value for which the 50\% of the recovered waveforms have bigger cross correlation.
%Coloured lines refer WhiteNoiseBurst with bandwidth [250,350] (black), [500,600] (red) and [1000,1100] (green) Hz.}}}
%\label{Fig. WNB 0d5}
%\end{figure}

\subsection{Compact Binary Coalescence}
For completeness, 
we also tested the algorithm on a set of signals coming from the coalescence of binary black hole systems. The first detections of gravitational waves were generated by the merge of two black holes that are the most cataclysmic events in nature.
In such systems two black holes combine to form a single one, emitting a strong gravitational wave. In our simulations we consider binary black hole systems with single masses between 15 and 25 solar masses and uniform spin distribution between 0 and 0.9.
%
%{\color{blue} --BISOGNA METTERCI DUE PAROLE SULLO SPIN A LIVELLO DI DEFINIZIONE}{\color{red}PERCHE'? LO SPIN NON E' UNA COSA GENERALE?}
%{\color{blue} LO DIAMO PER SCONTATO}
%
These are the same waveforms used in \cite{Reed}, 
where the cWB performances on sky localization have already been reported. %, with an approximation of 3.5 post-Newtonian of 
%IMRPhenomB, as used in \cite{Reed}.
Waveforms have been injected up to 4 Gpc, using a uniform distribution in volume. 
We are not considering any cosmological evolution, so in the previous values we are referring to masses in the source frames and luminosity distance.
The reconstructed events have been collected in
bins of network SNR, each bin has width equal to 2. To be homogeneous with the results of the other waveforms we consider a SNR 
range from 8 to 38.
Results from these tests are reported in Fig. \ref{Fig. Caserresult 0d5 CBC}. 
The results are similar to the case of White Noise Bursts in Fig. \ref{Fig. Caseresult 0d5 WNB}.
Indeed the signals belonging to these two classes involve a greater number of pixels in the time-frequency domain. This increases the noise contribution on the waveform estimation, which makes more difficult to accurately reconstruct the complete waveform.
%\begin{figure}[!hbt]
%\begin{center}
%%\begin{tabular}{cc}
%%\includegraphics[width=42mm]{CBC_hp_0d5.png} &
%%\includegraphics[width=42mm]{CBC_hc_0d5.png}
%%\end{tabular}
%\includegraphics[width=90mm]{CBC_hphc_0d5.png}
%\end{center}
%\caption{\small{\textit{CF50\% of the plus (left) and cross (right) polarization for injections uniformly distributed 
%in the sky as a function of the recovered SNR. Waveforms are generated by binary black holes coalescence with single component
%masses between 15 and 25 solar masses.
%Coloured lines refer to cases labelled in the text as \textbf{a)} (red), \textbf{b)} (green), \textbf{c)} (orange), \textbf{d)} (blue).}}}
%\label{Fig. CBC 0d5}
%\end{figure}

%\section{Results}

\section{Conclusion}

We have tested a new algorithm for the inverse solution as from the information given by a non-template search and reconstructed the original waveform polarization. This approach is the first attempt to use and verify the accuracy of the tool. 
Assuming the detectors at the design sensitivity and Gaussian noise, results show a reliable reconstruction of both plus $(+)$ and cross $(\times)$ polarizations, even for low injected values of SNR. \\Through the disentanglement of the 
various error contributions, we have shown that the main effect is due to the detector response. 
In fact, the correction caused by the sky position is mainly a time-shift of the signal proper time, but no significant distortion appears on the original waveform, as seen in Fig. \ref{Fig. Caseresult}. \\
We are working to obtain further improvements and refinement of the reconstruction by reducing the contribution of the noise in the detector response estimation. 
In future works we propose to verify if real detector noise gives comparing results and to study the effect of the matrix deficiency, for instance applying the approach proposed in \cite{Rakhmanov}. 
This tool does not rely on a specific pipeline, hence, it can be applied to any algorithm that provides information of the detector response and sky location. 
It would be interesting to apply this approach to detected events of Advanced LIGO and Virgo in the future observational runs. For events detected with template searches, it would be straightforward to compare the polarizations
obtained with this method with the ones given by the template. This would allow to check the performances of this
approach with the best matching, but also can give hints on possible variations of the real signal from the template itself.
For events detected with un-modeled algorithms, reconstructing the polarizations would be the first step to understand the generating process, other than the astrophysical progenitor.

\section{Acknowledgments}
The authors gratefully acknowledge the support of the research by the Italian Istituto Nazionale di Fisica
Nucleare and the Max-Planck-Society, and the State of Niedersachsen/Germany. They also gratefully 
acknowledge the support of L'Or{\'e}al Italia For Women in Science.


\begin{thebibliography}{99}
\bibitem{ein} A. Einstein, Sitzungsber. K. Preuss. Akad. Wiss. \textbf{1} 688, 1916.
\bibitem{ein1} A. Einstein, Sitzungsber. K. Preuss. Akad. Wiss. \textbf{1} 154, 1918.
\bibitem{hulse} R. A. Hulse and J. H. Taylor, \emph{Astrophys. J.} \text{195} L51, 1975.
\bibitem{taylor}  H. Taylor and J. M. Weisberg, \emph{Astrophys. J.}, \text{253}, 908 (1982)
%\bibitem{dooley} K. Dooley et al., \emph{J. Phys. Conf. Ser.}, \textbf{610(1)} 012012 {2015}.
\bibitem{LIGO} J. Aasi et al, \emph{Class. Quantum Grav.} \textbf{32}, 11, 2015
\bibitem{VIRGO} T Accadia, et al., \emph{Journal of Instrumentation} \textbf{7} P03012, 2012
\bibitem{S6} J. Abadie et al., \emph{Phys. Rev. D} \textbf{85} 122007, 2012.
\bibitem{S6CBC} J. Abadie et al., \emph{Phys. Rev. D} \textbf{85} 082002, 2012.
\bibitem{S6CW} J. Abadie et al., \emph{Phys. Rev. D} \textbf{90} 062010, 2014.
\bibitem{S6Stoch} J. Aasi et al., \emph{Phys. Rev. Lett} \textbf{113} 231101, 2014.
\bibitem{geo} Grote H et al., \emph{Class. Quantum Grav.} \textbf{27} 084003, 2010.
\bibitem{AdL} J. Abadie et. al, \emph{Class. Quantum Grav.} \textbf{32} 074001, 2015.
\bibitem{AdV} F. Acernese, et. al, %Advanced Virgo: a 2nd generation interferometric gravitational wave detector, 
\emph{Class. Quantum Grav.} \textbf{32} 024001, 2015.
\bibitem{Harry} G. M. Harry. \emph{Class. Quant. Grav.}, \textbf{27}  084006  (2010).
\bibitem{VC} The Virgo Collaboration, \emph{Advanced Virgo Technical Design Report}, 2012. [VIR-0128A-12].
\bibitem{Kagra} Y. Aso et al., 
%Y. Michimura, K. Somiya, M. Ando, O. Miyakawa, T. Sekiguchi, D. Tatsumi, and H. Yamamoto, Interferometer design of the KAGRA gravitational wave detector, 
\emph{Phys. Rev. D} \textbf{88} 043007, 2013.%, https://journals.aps.org/prd/abstract/10.1103/PhysRevD.88.043007 
\bibitem{Ken} K. Somiya,  \emph{Class. Quantum Grav.} \textbf{29} 124007, 2012.
\bibitem{India} B. Iyer, et al., LIGO-India Tech. rep. (2011), https://dcc.ligo.org/LIGO-M1100296/public 
\bibitem{event} B. P. Abbott et al., \emph{PRL} \textbf{116}, 061102, 2016.
\bibitem{Fairhust} S. Fairhust, \emph{Classical Quantum Gravity} \textbf{28}, 105021, 2011
\bibitem{burstPaper}  B. P. Abbott et al., \emph{PRD}  \textbf{93}, 122004, 2016.
\bibitem{cbcPaper}  B. P. Abbott et al., \emph{PRD}  \textbf{93}, 122003, 2016.
\bibitem{cWB2008} S. Klimenko et al., \emph{Classical Quantum Gravity} \textbf{25}, 114029, 2008.
\bibitem{cWB2015} S. Klimenko et al., Phys. Rev. D 93, 043007, 2016.
%\bibitem{astroevent} B. Abbott et al., \emph{Astrophys.J.} \textbf{818}  2, L22, 2016.
\bibitem{astroevent} B. P. Abbott et al., \emph{Phys. Rev. Lett.} \textbf{116}  221101, 2016.
\bibitem{bbcevent} B. P. Abbott et al., \emph{Phys. Rev. Lett.} \textbf{116} 241102, 2016.
\bibitem{BoxingDay} B. P. Abbott et al., \emph{Phys.Rev.Lett.} \textbf{116}, 241103, 2016.
\bibitem{GW170104} B. P. Abbott et al., \emph{Phys.Rev.Lett.} \textbf{118}, 221101, 2017.
\bibitem{GW170806} B. P. Abbott et al., arXiv:1711.05578
\bibitem{BBH-Mergers} B. P. Abbott et al., \emph{Phys.Rev.} \textbf{X6} no.4, 041015, 2016
\bibitem{GW170814} B. P. Abbott et al., \emph{Phys.Rev.Lett.} \textbf{119}, 141101, 2017
\bibitem{GW170817} B. P. Abbott et al., \emph{Phys.Rev.Lett.} \textbf{119}, 161101, 2017
\bibitem{GWGRB} B. P. Abbott et al., \emph{Astrophys.J.} \textbf{848}, no. 2, L13, 2017
\bibitem{Hubble} B. P. Abbott et al., \emph{Nature}, 10.0138/nature24471, 2017
\bibitem{AstrophImplic} B.P. Abbott (Caltech) et al., \emph{Astrophys.J.} \textbf{818} no.2, L22, 2016
\bibitem{S5} J. Abadie et al., \emph{Phys. Rev. D} \textbf{81} 102001, 2010.
\bibitem{cha} S. Chatterji, A. Lazzarini, L. Stein, P. Sutton, A. Searle and M. Tinto, \emph{Phys. Rev. D} \textbf{74} 082005, 2006.
\bibitem{and} W. G. Anderson, P. R. Brady, J. D. E. Creighton and E. E. Flanagan, \emph{Phys. Rev. D} \textbf{63} 042003, 2001.
\bibitem{wen} L. Wen and B. Schutz, \emph{Class. Quantum Grav.} \textbf{22} S1321, 2005. 
\bibitem{laura} L. Cadonati, \emph{Class. Quantum Grav.} \textbf{21} S1695 2004.
\bibitem{BW} N. Cornish and T. Littenberg,, \emph{Class. Quantum Grav.} \textbf{32}, 13  2015.
\bibitem{sch} B. S. Sathyaprakash and B. F. Schutz, \emph{Living Reviews in Relativity}, \textbf{12}(2), 2009.
\bibitem{Sergey}  S. Klimenko et al, \emph{Phys.Rev.D} \textbf{72}, 122002, 2005.
\bibitem{PRD2011} S. Klimenko et al, \emph{Phys. Rev. D} \textbf{83}, 102001, 2011.
\bibitem{Reed} R. Essick et al., The Astrophysical Journal, 800, 2
\bibitem{Rakhmanov} M. Rakhmanov \emph{Class. Quantum Grav.} \textbf{23}, S673-S685, 2006.
\bibitem{LIGO psd} https://dcc.ligo.org/LIGO-T0900288/public.
\bibitem{Virgo psd} https://tds.ego-gw.it/ql/?c=6589.
\bibitem{O1} B. P. Abbott et al., \emph{Phys. Rev. D} \textbf{95} 042003, 2017.
%\bibitem{lowlatency} M. Drago et al, in preparation
%\bibitem{Healpix} http://healpix.jpl.nasa.gov/.
%\bibitem{LALinference} J. Veitch et al., \emph{Phys. Rev. D} \textbf{91}, 042003 (2015)
\end{thebibliography}
\end{document}